\documentclass[twocolumn,showpacs,amsmath,amssymb,superscriptaddress,eqsecnum,prb]{revtex4}

\usepackage{graphicx}
\usepackage{dcolumn}
\usepackage{bm}

\usepackage{color}
\definecolor{red}{rgb}{0.8,0.0,0.0}

\newcommand{\aver}[1]{\langle #1 \rangle}
\DeclareMathOperator{\Tr}{Tr}

\begin{document}

\title{Frequency-Dependent Current Noise through Quantum-Dot Spin Valves}

\author{Matthias Braun}
\affiliation{Institut f\"ur Theoretische Physik III, Ruhr-Universit\"at Bochum, 44780 Bochum, Germany}
\author{J\"urgen K\"onig}
\affiliation{Institut f\"ur Theoretische Physik III, Ruhr-Universit\"at Bochum, 44780 Bochum, Germany}
\author{Jan Martinek}
\affiliation{Institute for Materials Research, Tohoku University, Sendai 980-8577, Japan}
\affiliation{Institut f\"ur Theoretische Festk\"orperphysik, Universit\"at Karlsruhe, 76128 Karlsruhe, Germany}
\affiliation{Institute of Molecular Physics, Polish Academy of Science, 60-179 Pozna\'n, Poland}

\date{\today}

\begin{abstract}
We study frequency-dependent current noise through a
single-level quantum dot connected to ferromagnetic leads
with non-collinear magnetization. We propose to use the 
frequency-dependent Fano factor as a tool to detect
single-spin dynamics in the quantum dot.
Spin precession due to an external magnetic and/or a many-body
exchange field affects the Fano factor of the system in two
ways. First, the tendency towards spin-selective bunching of
the transmitted electrons is suppressed, which gives rise to
a reduction of the low-frequency noise. Second, the noise
spectrum displays a resonance at the Larmor frequency,
whose lineshape depends on the relative angle of the leads'
magnetizations.
\end{abstract}

\maketitle

\section{\label{Introduction}Introduction}
The measurement of current noise reveals additional
information about mesoscopic conductors that is not contained
in the average current.\cite{review,review2} Current noise through
quantum dots exposes the strongly-correlated character of
charge transport due to Coulomb interaction, giving rise to
phenomena such as positive cross correlations,\cite{cottet}
and sub- or super-Poissonian Fano factors.\cite{superpoissonian1,superpoissonian2}
This is one motivation for the extensive
theoretical\cite{korotkov,hershfield,hanke,thielmann,Lue} and
experimental\cite{experiment,haug1,haug2} study of zero- and
finite-frequency noise of the current through quantum dots.
Furthermore, the finite-frequency noise provides a direct
access to the internal dynamics of the system such as
coherent oscillations in double-dot
structures,\cite{gurvitz2,bingdong,Sun,Kieslich} quantum-shuttle
resonances,\cite{flindt} transport through a dot with a
precessing magnetic moment,\cite{precessingspin} or back
action of a detector to the
system.\cite{backaction1,backaction2,backaction3}

\begin{figure}[ht]
\includegraphics[width=0.9\columnwidth]{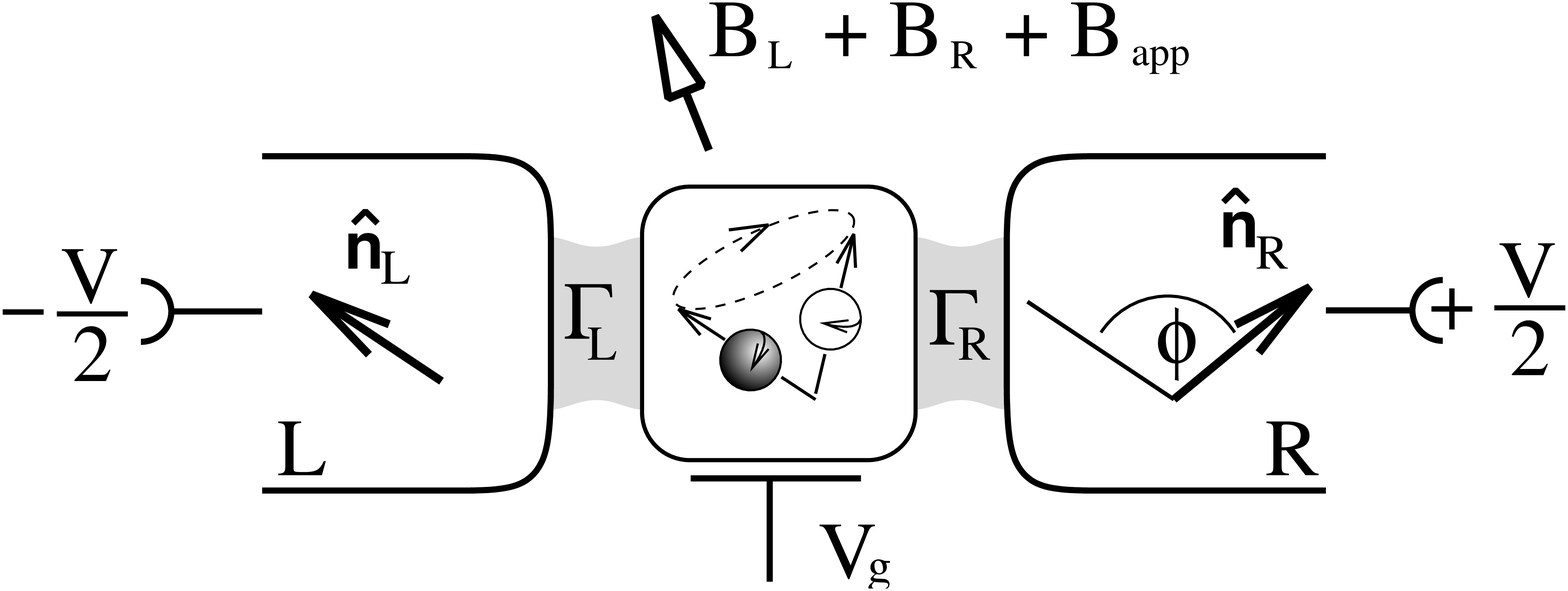}
\caption{\label{fig:system}A quantum dot contacted by
ferromagnetic leads with non-collinear magnetizations.
Electrons polarized along the source (left) lead enter the dot. During their
stay on the dot, the spins precess in the many-body exchange field
$\bm B_{\rm L}+\bm B_{\rm R}$, which arises from the tunnel coupling to the left and right lead, and an applied magnetic field $\bm B_{\rm app}$. Due to
magnetoresistance effects this precession modulates the
tunnel-out probability to the drain (right) lead, giving rise to a signal in the power spectrum of the current noise.}
\end{figure}

In this paper, we investigate the transport through a
single-level quantum dot connected to ferromagnetic
leads with non-collinear magnetizations in the
limit of weak dot-lead coupling, see Fig.~\ref{fig:system}.
Recent experimental approaches to contact a
 quantum dot to ferromagnetic leads involve metallic islands,
\cite{ono,wees} granular systems, \cite{grain_experiment,
granular} carbon nanotubes \cite{cnt,schoenenberger}
as well as single molecules \cite{ralph2}
or self-assembled quantum dots.\cite{QD-exp,ralph}
Quantum-dot spin-valve structures are interesting,
since the presence of both a finite spin polarization in the leads
and an applied bias voltage induces, for a non-parallel
alignment of the lead magnetization directions, an
non-equilibrium spin on the quantum dot. The magnitude and
direction of the quantum-dot spin is determined by
the interplay of two processes: non-equilibrium spin
accumulation due to spin injection from the leads,
and spin precession due to an exchange field generated
by the tunnel coupling to spin-polarized leads\cite{braun1}
or due to an externally applied
magnetic field.\cite{braun2} The resulting average
quantum-dot spin affects the $dc-$conductance of the device.

While the time-averaged current is sensitive
to the time-averaged dot spin, the time-resolved
dynamics of the dot spin is provided by the power
spectrum of the current noise.
It will show a signature at the frequency that is associated with
the precession of the quantum-dot spin due to the sum of exchange
and external magnetic field. This can be understood by looking at
the tunneling-out current to the drain (right) lead as a function of
the time after the quantum-dot electron had tunneled in from the
source (left) lead. The spin of the incoming electron, defined by
the source-lead magnetization direction, precesses about the sum of
exchange and external magnetic field as long as it stays in the dot.
Since the tunneling-out rate depends on the relative orientation of
the quantum-dot spin to the drain-lead magnetization direction,
the spin precession leads to a periodic oscillation of the tunneling-out
probability. The period of the oscillation is defined by the inverse precession
frequency, and the phase is given by the relative orientation of
the source- and drain-lead magnetization direction.
As a consequence, the signature in the power spectrum of the current
noise at the Larmor frequency gradually changes from a peak to a dip
as a function of angle between source- and drain-lead magnetization.

Also the zero-frequency part of the current-noise power spectrum is
affected by the internal dynamics of the quantum-dot spin.
 By coupling a quantum dot to spin-polarized electrodes, the
dwell time of the electrons in the dot becomes spin dependent.
It is known\cite{Bulka1999, cottet} that this spin dependence of the dwell
times yields a bunching of the transferred
electrons, which leads to an increase of the shot noise.
A precession of the quantum-dot spin due to exchange and external
magnetic field weakens the tendency towards bunching,
leading to a reduction of the
low-frequency noise.

The aim of this paper is to perform a systematic study of the
frequency-dependent current noise of a quantum-dot spin valve in the
limit of weak dot-lead coupling in order to illustrate the effects
formulated above. In Sec.~\ref{System} we define the model of a
quantum-dot spin valve, as shown in Fig.~\ref{fig:system}.
In Sec.~\ref{Technique}, we extend a previously-developed
diagrammatic real-time technique\cite{diagrams} to evaluate
frequency-dependent current noise, as it has been similarly done
for metallic (non-magnetic) single-electron
transistors.\cite{johansson,backaction1} The results for the
quantum-dot spin valve are discussed in Sec.~\ref{Results},
followed by the Conclusions in Sec.~\ref{conclusions}.

\section{\label{System}Model System}
The Hamiltonian for the quantum-dot spin valve, i.e., a
quantum dot coupled to ferromagnetic leads, is given by the sum
\begin{eqnarray}
  H= H_{\rm L} + H_{\rm R} + H_{\rm D} + H_{\rm T} \,.
\end{eqnarray}
The single-level quantum dot is modeled by an Anderson impurity,
\begin{eqnarray}\label{Hdot}
  H_{\rm D}=\sum_{\sigma=\uparrow\downarrow}\varepsilon_\sigma
  c^{\dag}_{\sigma}c_{\sigma} + U\,n_{\uparrow}n_{\downarrow}\,,
\end{eqnarray}
where $c^{\dag}_{\sigma}$ and $c_{\sigma}$ are the fermion
creation and annihilation operators of the dot electrons, and
$n_{\sigma}=c^{\dag}_{\sigma}c_{\sigma}$. The single-particle level
at the energy $\varepsilon$, measured relative
to the equilibrium Fermi energy of the leads, may be split due to
an external magnetic field,
$\varepsilon_\uparrow = \varepsilon +\Delta/2$ and
$\varepsilon_\downarrow = \varepsilon -\Delta/2$ with Zeeman energy
$\Delta =g \mu_{\rm B} B_{\rm ext}$.
Double occupancy of the dot costs the charging energy
$U\gg k_{\rm B}T$.

The ferromagnetic leads $(r=\rm L / R)$ are treated as reservoirs of
non-interacting fermions,
\begin{eqnarray}
  H_r = \sum_{k,\alpha=\pm} \varepsilon^{\,}_{k\alpha} a^{\dag}_{rk\alpha}
  a^{\,}_{rk\alpha}\, .
\end{eqnarray}
By choosing the quantization axis of each lead parallel to their
direction of magnetization ${\bf \hat{n}}_{r}$, the property of
ferromagnetism can be included by assuming different
density of states $\xi_{\alpha}$ for majority $(\alpha=+)$ and
minority $(\alpha=-)$ electrons.
An applied bias voltage is incorporated by a symmetric shift of the
chemical potential by $\mu_{\rm L/R}=\pm eV/2$ in the left and right lead, which
enter the Fermi functions $f_r(E) = f(E - \mu_r)$.

The magnetization directions of the left and right lead and the
external magnetic field are, in general, non-collinear, i.e., in the
Hamiltonians for the three subsystems we have chosen different spin
quantization axes. To describe spin-conserving tunneling, one must
include $SU(2)$ rotation matrices $U_{\alpha\sigma}^{r}$ in
the tunneling Hamiltonian
\begin{eqnarray}
  \label{Htunnel}
  H_{\rm T}= \sum_{r,k,\sigma\alpha}t_{r}
  a^{\dag}_{rk\alpha}U_{\alpha\sigma}^{r}c^{}_{\sigma} +
  \,\text{H.c.}\,.
\end{eqnarray}
For simplicity we use leads with energy-independent
density of states $\xi_{\alpha}$ and barriers with
energy-independent tunnel amplitudes $t_r$.
With these assumptions, the degree of lead polarization
$p=(\xi_{+} -\xi_{-})
  /(\xi_{+}+\xi_{-})$ as well
as the coupling constants $\Gamma_r=\sum_{\alpha=\pm}2\pi|
t_r/\sqrt{2}|^2 \xi_{\alpha}$ do not depend on energy.

\section{\label{Technique}Diagrammatic Technique}
The dynamics of the
quantum-dot spin valve is determined by the time evolution
of the total density matrix. Since the leads are modeled by
non-interacting fermions, which always stay in equilibrium,
we can integrate out
the degrees of freedom in the leads, and only need to consider
the time evolution of the reduced density matrix ${\bm \rho}(t)$
of the quantum dot, which contains the information about both
the charge and spin state of latter.
In the following three subsections, we formulate the derivation
for the stationary density matrix, the $dc-$current and the finite-frequency
current-current correlation function.
Afterwards, in Sec.~\ref{Sequential}, we specify the obtained formulas
for the limit of weak dot-lead coupling, i.e. we perfrom a systematic
lowest-order perturbation expansion in the tunnel coupling strength
$\Gamma=\Gamma_{\rm L}+\Gamma_{\rm R}$.

\subsection{\label{densitymatrix}Density matrix}

\begin{figure}[h!]
\includegraphics[width=0.9\columnwidth]{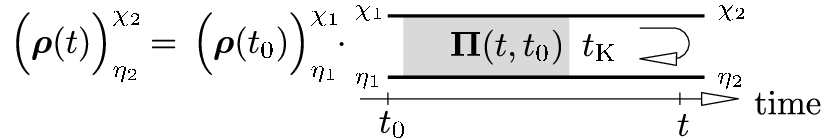}
\caption{\label{propagator1}
The density matrix evolves in time with the propagator
$\bm \Pi$, which is a tensor of rank four.}
\end{figure}

The quantum-statistical average of the charge and spin on the quantum dot at
time $t$ is encoded in the reduced density matrix ${\bm \rho}(t)$.
Its time evolution is governed by the propagator ${\bm \Pi}(t,t_0)$,
\begin{eqnarray}\label{prop0}
  {\bm \rho} (t) = {\bm \Pi}(t,t_0) \, \cdot {\bm \rho}(t_0) \,.
\end{eqnarray}
 Since $\bm \rho$ is a matrix, the propagator $\bm \Pi)$ must be 
a tensor of rank four. A diagrammatic representation of this equation (see also
Ref.~\onlinecite{diagrams}) is depicted in Fig.~\ref{propagator1}.
The upper/lower horizontal line represents the propagation of the individual
dot states forward/backward in (real) time, i.e. along a Keldysh
time contour $t_{\rm K}$.

\begin{figure}[h!]
\includegraphics[width=0.9\columnwidth]{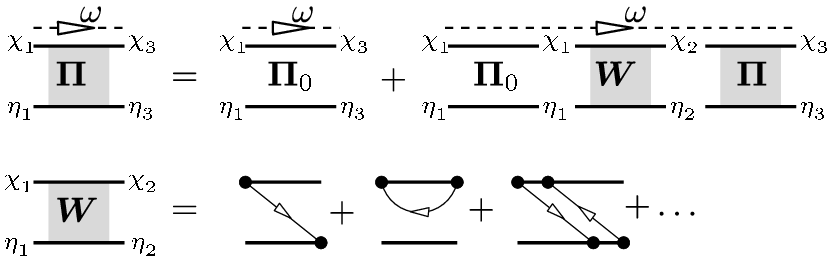}
\caption{\label{dyson3}
  Diagrammatical representation of the Dyson equation for 
  the propagator. The self energy $\bm W$ sums up all 
  irreducible tunnel diagrams. With $\bm W(\omega)$, we 
  label the self energy, together with the parallel 
  running frequency line $\omega$.}
\end{figure}

In order to find the stationary density matrix for a system, which is
described by a time-independent Hamiltonian, we consider the limit
$t_0 \rightarrow - \infty$.
There is some characteristic time after which the system loses the information
about its initial density matrix $\bm \rho_{\rm ini} =
\lim_{t_0\rightarrow -\infty} \bm \rho(t_0)$.
We can, therefore, choose without loss of generality
$\left( \bm \rho_{\rm ini} \right)^{\chi_1}_{\eta_1} = \delta_{\chi_1 , \chi_0}
\delta_{\eta_1 , \chi_0}$ with an arbitrarily-picked state $\chi_0$, to get
for the stationary (non-equilibrium) density matrix
\begin{eqnarray}\label{limit}
  {\bigl(\bm \rho_{\rm st}\bigl)}_{\eta_2^{}}^{\chi_2^{}} =
  \lim_{t_0\rightarrow-\infty}
      {\Pi(t-t_0)\vphantom{\bigl)}}_{\eta_2^{} \chi_0^{}}^{\chi_2^{} \chi_0^{}}
\, ,
\end{eqnarray}
independent of $\chi_0$.
 Here, for time-translation invariant systems, the propagator
$\bm \Pi(t,t_0)$ depends only on the difference of the time
arguments $(t-t_0)$.
For the following, it is convenient to express the propagator in frequency
representation $\bm \Pi(\omega)= \hbar^{-1}\int_{-\infty}^{0} dt\, \bm \Pi(-t)\,
\exp[i (\omega - i0^+) t]$.
It can be constructed by the Dyson equation
\begin{eqnarray}\label{propagator}
{\bm\Pi}(\omega)&=&{\bm\Pi}_0(\omega)+{\bm\Pi_0}(\omega){\bm W}(\omega){\bm\Pi}(\omega)\nonumber\\
&=&\left[{{\bm\Pi}_0}^{-1}(\omega)-{\bm W}(\omega)\right]^{-1}\,.
\end{eqnarray}
The full propagator ${\bm\Pi}(\omega)$
depends on the free propagator ${\bm\Pi}_0(\omega)$ and
the irreducible self energies $\bm W(\omega)$, which describes the
influence of tunneling events between the dot and the leads. The
Dyson equation is diagrammatically represented in Fig.~\ref{dyson3}.
The frequency argument of the Laplace transformation appears in this
diagrammatic language\cite{diagrams} as additional horizontal bosonic line
transporting energy $\hbar\omega$.

The free propagator (without tunneling) is given by
\begin{eqnarray}
  \label{freepropagator}
	{\Pi_0(\omega)}_{\eta_2^{} \eta_1^{}}^{\chi_2^{} \chi_1^{}}
	=\frac{i\delta_{\eta_1^{} \eta_2^{}}\delta_{\chi_1^{} \chi_2^{}}}{
	  \varepsilon_{\eta_1^{}}-\varepsilon_{\chi_1^{}}
	  -\hbar\omega+i0^+}\,,
\end{eqnarray}
where $\varepsilon_\chi$ ($\varepsilon_\eta$) is the energy of the
dot state $\chi$ ($\eta$). Tunneling between the dot and the leads
introduce the irreducible self energies $\bm W(\omega)$. We
calculate $\bm W(\omega)$ in a perturbation expansion in the tunnel
Hamiltonian Eq.~(\ref{Htunnel}). Each tunnel Hamiltonian generates
one vertex (filled circle), on the Keldysh time contour $t_{\rm K}$,
see Fig.~\ref{dyson3}. Since the leads are in equilibrium, their
non-interacting fermionic degrees of freedom can be integrated out. Thereby two
tunnel Hamiltonians each get contracted, symbolized by a line. Each
line is associated with one tunnel event, transferring one particle
and a frequency/energy from one vertex to the other. Therefore the
lines have a defined direction and bear one order of the coupling
constant $\Gamma=\Gamma_{\rm L}+\Gamma_{\rm R}$. We define the self
energy $\bm W(\omega)$ as the sum of all irreducible tunnel diagrams
(diagrams, which can not be cut at any real time, i.e. cut
vertically, without cutting one tunneling line).

 In Sec.~\ref{Sequential}, we will then restrict
our otherwise general calculation
to the lowest-order expansion in $\Gamma$, i.e. we will
include only diagrams with one tunnel line in  $\bm W(\omega)$.
A detailed description of how to calculate these lowest-order self
energies as well as example calculations of $\bm W$ for
the system under consideration can be found in
Ref.~\onlinecite{braun1}.

To solve for the stationary density matrix  ${\bm \rho}_{\rm st}$,
we rewrite the Dyson
Eq.~(\ref{propagator}) as $({\bm\Pi_0}(\omega)^{-1}-{\bm W}(\omega))
{\bm\Pi}(\omega)=\bm 1$, multiply both sides of
the equation with $\omega$, use
the final value theorem $\lim_{\omega\rightarrow0}(i\hbar\omega+0^+)\bm \Pi(\omega)=
\lim_{t\rightarrow\infty}\bm \Pi(t)$, similar as for Laplace transformations,
and employ Eq.~(\ref{limit}), to get the generalized master equation
\begin{eqnarray}\label{master}
\bm 0=\Bigl[\bm \Pi_0^{-1}(\omega =0)-{\bm W}(\omega =0)\Bigl]\,\,
{\bm \rho}_{\rm st}
\end{eqnarray}
together with the normalization condition $\Tr[\bm \rho_{\rm st}]=1$.

The structure of
Eq.~(\ref{master}) motivates the interpretation of the self energy
$\bm W(\omega=0)$ as generalized transition rates. However, the self energy
does not only describe real particle transfer between leads and dot,
but it also accounts for tunneling-induced renormalization effects.
It was shown in Refs.~\onlinecite{braun1,spincurrent,doubledot,Braig2005}, that these
level renormalization effects may affect even the lowest-order contribution
to the conductance. Therefore, a neglect of these renormalizations would
break the consistancy of the lowest-order expansion in the tunnel coupling
strength.\cite{bingdong,RudzinskiBarnas,gurvitz1}
Recently, the frequency-dependent current noise of a quantum-dot spin 
valve structure was discussed in Ref.~\onlinecite{gurvitznoise}, in 
the limit of infinite bias voltage, where these level renormalizations can 
be neglected. One of the main advantages of the approach presented here 
is, that a rigid systematic computation of the generalized transition rates 
is possible, which include all renormalization effects. Therefore our approach 
is valid for arbitrary bias voltages.

\subsection{\label{current}Current}
\begin{figure}[h!]
\includegraphics[width=0.9\columnwidth,angle=0]{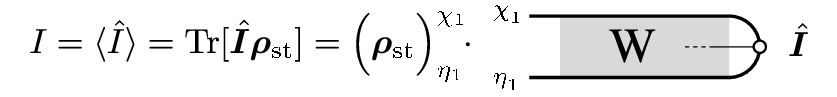}
\caption{\label{fig:current}
Diagrammatic representation of the current. By integrating out the
lead degrees of freedom, the current vertex (open circle) gets
contracted to one of the tunnel vertex in a self energy $\bm W(\omega=0)$.}
\end{figure}

The current through barrier $r={\rm L},{\rm R}$ is defined as the change of
charge $en_r= e\sum_{k\sigma} a^{\dag}_{rk\sigma}a^{\,}_{rk\sigma}$ in lead
$r$ due to tunneling, described by the operator
\begin{eqnarray}\label{eq:current}
\hat{I}_r
=e \frac{\partial n_r}{\partial t}
=\frac{e}{i\hbar}[n_r,H_{\rm T}]\,.
\end{eqnarray}
We define the operator for the current through the dot as
$\hat{I}=(\hat{I}_{\rm L}-\hat{I}_{\rm R})/2$.
Each term of the resulting current operator does contain
a product of a lead and a dot operator. By integrating out
the  lead degrees of freedom, the current vertex (open circle)
gets connected to a tunnel vertex by a contraction line as
depicted in Fig.~\ref{fig:current}. Thereby the tunnel vertex
can be either on the upper or lower time contour line.

To present a systematic way to calculate the current, we
can utilize the close similarity of the tunnel Hamiltonian
in Eq.~(\ref{Htunnel}) and the current operator in
Eq.~(\ref{eq:current}). Both differ only by the prefactor
$e/\hbar$ and  possibly by additional minus signs.

\begin{figure}[h!]
\includegraphics[width=0.9\columnwidth,angle=0]{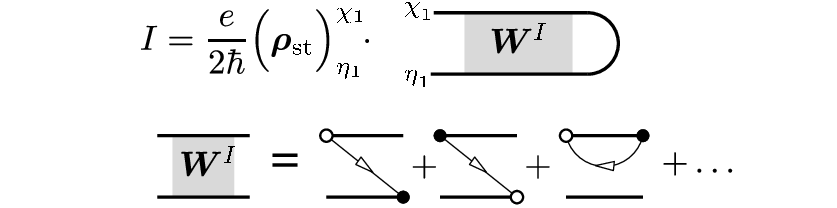}
\caption{\label{fig:currentb}
Reformulation of the current as function of $\bm W^I(\omega=0)$,
the self energy with one tunnel vertex replaced by a
current vertex.}
\end{figure}

Following the work of Thielmann {\it et al.},\cite{thielmann} we define
the object ${W^I}^{\chi_1 \chi}_{\eta_1 \eta}$ as the sums of all
possible realizations of replacing one tunnel vertex
(filled circle) by a current vertex (open circle) in
the self energy ${W}^{\chi_1 \chi}_{\eta_1 \eta}$, compare
Fig.~\ref{fig:currentb}.
In technical terms, this means that each diagram is multiplied
by a prefactor, determined by the position of the current vertex
inside the diagram. If the current vertex is on the upper (lower)
Keldysh time branch, and describes a particle tunneling into the
right (left) lead or out of the left lead (right), multiply the
diagram by $+1/2$, otherwise by $-1/2$.  For clarity, we keep the
factor $e/\hbar$ separate.
For the detailed technical procedure of the replacement
as well as the rules to construct and calculate the
self energies, we refer to Ref.~\onlinecite{thielmann}.
The average of one current operator, i.e. the $dc-$current flowing
through the system is then given by
\begin{eqnarray}
I=\aver{\hat{I}}= \frac{e}{2 \hbar}  \Tr[{\bm W}^I(\omega=0)\,{\bm \rho}_{\rm st}]\,.
\end{eqnarray}
The trace selects the diagonal matrix elements,
which regards that the Keldysh line must be closed
at the end of the diagram, see Fig.~\ref{fig:currentb},
requiring that the dot state of the upper and lower time branch match.

To see, that the diagrams in Fig. \ref{fig:current} and
Fig. \ref{fig:currentb} are equal, one
must consider, that all diagrams, where the 
rightmost vertex is a tunnel vertex will cancel each 
other when performing the trace. This happens, since by moving the
rightmost tunnel vertex from the upper (lower) to the
lower (upper) Keldysh time line, the diagram aquires
only a minus sign.\cite{diagrams}

\subsection{\label{noise}Current-current correlation}

We define the frequency-dependent noise as the Fourier transform of
$S(t)=\aver{\hat{I}(t)\hat{I}(0)} +\aver{\hat{I}(0)\hat{I}(t)}
-2\aver{\hat{I}}^2$, which can be written as
\begin{eqnarray}\label{noisedefinition}
S(\omega)&=&\int^{\infty}_0 \!\!\!\!\!dt
\Bigl( \aver{\hat{I}(t)\hat{I}(0)}+\aver{\hat{I}(0)\hat{I}(t)}\Bigl)
\Bigl( e^{-i \omega t} + e^{+i \omega t}\Bigl)\nonumber\\
  && -4\pi\,\delta(\omega)\,\aver{\hat{I}}^2\,.
\end{eqnarray}
We restrict our discussion to the above defined symmetrized and,
therefore, real noise since it can be measured by a classical
detector.\cite{quantummeasuremnet} The unsymmetrized noise would have an
additional complex component, describing absorption and emission
processes,\cite{assymmetric} that depend on the specifics of the detector.

Since the dot-lead interface capacitances are much less sensitive to
the contact geometry than the tunnel couplings $\Gamma_{\rm L/R}$,
we assume an equal capacitance of the left and right interface, while
still allowing for different tunnel-coupling strengths. Following the
Ramon-Shockley\cite{theorem} theorem, we have then to define current operator also symmetrized with respect to the left and right
interface as already done in Sec.~\ref{current}.

\begin{figure}[h!]
\includegraphics[width=0.9\columnwidth]{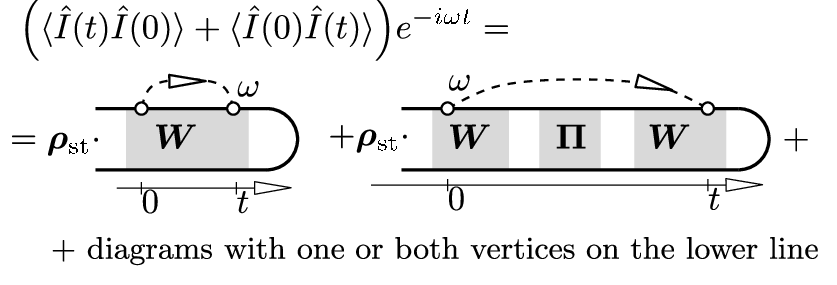}
\caption{\label{WIIWIPWI}
Regrouping of the noise expansion by introducing
the irreducible self energy $\bm W$, and the propagator $\bm \Pi(\omega)$.}
\end{figure}

The diagrammatic calculation of the current-current correlation
function is now straightforward. Instead of replacing one tunnel
vertex by a current vertex on the Keldysh time contour, as for the
average current, one must replace two vertices.
The additional frequency $\omega$ of the Fourier transformation in
Eq.~(\ref{noisedefinition}) can be incorporated in the diagrams as an
additional bosonic energy line (dashed) running from  $t$ to $0$,
i.e. between the two current vertices.\cite{johansson} This line
must not be confused with a tunnel line, since it only transfers
energy $\hbar \omega$, and no particle. By introducing the self
energy $\bm W$ all diagrams of the current-current correlation
function can be grouped in two different
classes\cite{johansson,thielmann} as shown in Fig.~\ref{WIIWIPWI}.
Either both current vertices are incorporated in the same irreducible block
diagram, or into two different ones that are separated by the propagator
$\bm \Pi(\omega)$.

The order of the 
current operator on the Keldysh contour is determined by its
ordering in the correlator,  so the current operator at time $0$ lies
on the upper branch for $\aver{\hat{I}(t)\hat{I}(0)}$ and on the lower branch
for $\aver{\hat{I}(0)\hat{I}(t)}$. Since in Eq.~(\ref{noisedefinition}) we
defined the noise symmetrized with respect to the operator ordering,
we just allow every combination of current vertex replacements in
the $\bm W$'s. This includes also diagrams where one or both vertices 
are located on the lower time contour (this type of diagrams are not
explicitly drawn in Fig.~\ref{WIIWIPWI}).

By including the current vertices and the frequency line in the self
energies, three variants of the self energy ${\bm W}$ are
generated. The objects  ${\bm W}^I_{>}(\omega)$ and ${\bm W}^I_{<}
(\omega)$ are the sum of all irreducible diagrams, where one tunnel
vertex is replaced  by a current vertex in any topological different
way. The subindex $>(<)$ indicates, that the frequency line connected to
the current vertex leaves or enters the diagram to the right (left) side.
In the zero-frequency limit, the two objects become equal
${\bm W}^I_{>}(\omega=0)={\bm W}^I_{<}(\omega=0)\equiv {\bm W}^I$.

\begin{figure}[h!]
\includegraphics[width=0.9\columnwidth]{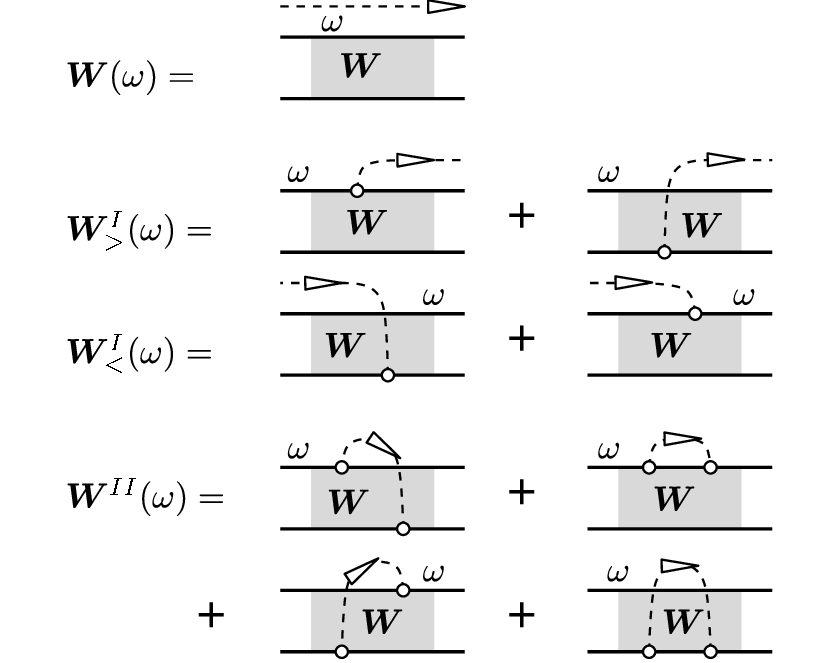}
\caption{\label{partsofdiagrams}
Different variations of the self energy $\bm W$.}
\end{figure}

The third object ${\bm  W}^{II}(\omega)$ sums irreducible diagrams with
two tunnel vertices replaced each by a current vertex in any
topological different way. The current vertices are connected by the
frequency line $\omega$. The diagrammatical picture of the objects
${\bm W}(\omega)$,
${\bm  W}^{II}(\omega)$, ${\bm W}^I_{>}(\omega)$, and
${\bm W}^I_{<}(\omega)$ are shown in Fig.~\ref{partsofdiagrams}.

With these definitions the diagrams for the frequency-dependent 
noise in Fig.~\ref{WIIWIPWI} can be directly translated into the 
formula
\begin{eqnarray}\label{noise2}
  S(\omega) &=&
  \frac{e^2}{2\hbar} \Tr[{\bm  W}^{II}(\omega){\bm \rho}_{\rm st} 
    +{\bm W}^I_{<}(\omega)\,{\bm \Pi}(\omega)\, 
    {\bm W}^I_>(\omega) {\bm \rho}_{\rm st}]
  \nonumber \\
  && - 2\pi \delta(\omega) \,\aver{\hat{I}}^2 +(\omega \rightarrow -\omega )\,.
\end{eqnarray}
We remark that the first line in Eq.~(\ref{noise2}) diverges as 
$\omega\rightarrow 0$.
While the $\bm W$'s are regular for $\omega\rightarrow 0$,
the propagator ${\bm \Pi}(\omega)$ goes as $i / (-\hbar\omega + i 0^+)$ 
times $\bm \Pi(t\rightarrow\infty)$, which is related to 
$ {\bm \rho}_{\rm st}$ via Eq.~(\ref{limit}).
In the limit $\omega\rightarrow 0$ the propagator therefore yields both 
a delta function $\delta (\omega)$ and a $1/\omega$  divergence.
For the full expression of the noise, 
these divergences are canceled by the delta-function term 
in the second line of Eq.~(\ref{noise2}) and by the terms with 
$\omega \rightarrow -\omega$, respectively.
As a consequence, $S(\omega)$ remains regular also in the limit
$\omega\rightarrow 0$.

\subsection{\label{Sequential}Low-frequency noise in the sequential-tunnel
limit 
}

Equation~(\ref{noise2}) is the general expression for the frequency-dependent 
current noise. In the following paper, we consider only the limit of weak 
dot-lead tunnel coupling, $\Gamma \ll k_{\rm B} T$, and therefore include only 
diagrams with at most one tunnel line in the $\bm W$'s.
However, this procedure is not a consistent expansion scheme for the noise 
$S(\omega)$ itself.  
By expanding the $\bm W$'s up to linear order in $\Gamma$, the result of 
Eq.~(\ref{noise2}) is the consistent noise linear in $\Gamma$ plus some 
higher-order contributions proportional to $\Gamma^2$. 
Since co-tunnel processes
also give rise to quadratic contributions, we have to discard these terms as
long as we neglect the quadratic cotunnel contributions of $\bm W$.
If one is interested in the noise up to second order in $\Gamma$,
then these higher-order terms generated by lower-order $\bm W$'s are of
course an essential part of the result.\cite{thielmann2}

Further, we are looking for signatures of the internal charge and spin 
dynamics of the quantum-dot in the frequency-dependent current noise. 
Therefore - if we neglect external magnetic fields at this point - 
we concentrate on frequencies that are at most of the same order
of the tunnel coupling $\Gamma$.
If we limit the range in which we want to calculate the current
noise to $\hbar \omega \lesssim \Gamma$, we can neglect the 
frequency dependence of the $\bm W$'s. Each correction of the $\bm W$'s 
would scale at least with $\omega\Gamma\approx \Gamma^2$, making them 
as important as the neglected co-tunnel processes.

The neglect of the terms in $\bm W$ which are at least linear 
in frequency has two main advantages.
First, it considerably simplifies the calculation of the $\bm W$'s. 
Second, it automatically removes the quadratic parts of the noise, 
so Eq.~(\ref{noise2}) gives a result consistent in linear order in $\Gamma$.
In this low-frequency limit, the noise can then be written as
\begin{eqnarray}\label{smallw}
S(\omega)&=&
\frac{e^2}{\hbar} \Tr
[{\bm  W}^{II}{\bm \rho}_{\rm st}+{\bm W}^I\,
\Bigl[{\bm \Pi}_0^{-1}(\omega)-{\bm W}\Bigl]^{-1}
\, {\bm W}^I{\bm \rho}_{\rm st}]\nonumber \\
  && - 2\pi \delta(\omega) \,\aver{\hat{I}}^2 +(\omega \rightarrow -\omega )
\end{eqnarray}
where
$\bm W^I\equiv{\bm W}^I_{>}(\omega=0)={\bm W}^I_{<}(\omega=0)$, ${\bm W}\equiv
{\bm W}(\omega=0)$, and ${\bm W}^{II}\equiv{\bm W}^{II}(\omega=0)$.
 This means, that the bosonic frequency lines $\omega$ in the
diagrams as shown in Fig.~\ref{partsofdiagrams} can
be neglected. The only remaining frequency-dependent part is the
free propagator ${\bm \Pi}_0(\omega)$.

This formalism, of course, reproduces the noise spectrum of a 
single-level quantum dot connected to normal leads as known from
literature.\cite{review} If one can approximate the Fermi functions
by one or zero only, i.e. if the dot levels are away from the Fermi
edges of the leads the Fano factor shows a Lorentzian dependence on 
the noise frequency $\omega$
\begin{eqnarray}
\label{pequal0ffnoise1}
F(\omega) & \equiv &\frac{S(\omega)}{2 e I} = \frac{1}{2}
+\frac{(2\Gamma_{\rm L}-\Gamma_{\rm R})^2}{(2\Gamma_{\rm L}+\Gamma_{\rm R})^2 + (\hbar\omega)^2}\qquad
\end{eqnarray}
for a bias voltage allowing only an empty or singly-occupied dot, and
\begin{eqnarray}
\label{pequal0ffnoise}
F(\omega) & = & \frac{1}{2}
+\frac{(\Gamma_{\rm L}-\Gamma_{\rm R})^2}{(\Gamma_{\rm L}+\Gamma_{\rm R})^2 + (\hbar\omega)^2}
\end{eqnarray}
for higher bias voltages, when double occupation is also allowed.

\subsection{\label{Technical}Technical summary}
The technical scheme for calculating the zero- and
low-frequency current noise is the following:
First the objects $\bm W$, $\bm W^I$ and $\bm W^{II}$
must be calculated in the $\omega=0$ limit, using the
diagrammatic approach, see Ref.~\onlinecite{braun1}.

 In the next step, we calculate the reduced density
matrix $\bm \rho$ of a single-level quantum dot, which
is a $4\times 4$ matrix,
\begin{eqnarray}
  \label{dotdm}
  {\bm \rho}= \left(
  \begin{array}{cccc}
    \rho_0^0 &     0 &     0 &     0 \\
    0     & \rho_{\uparrow}^{\uparrow} & \,\rho^{\uparrow}_{\downarrow} &     0 \\
    0     & \,\rho^{\downarrow}_{\uparrow} & \rho_{\downarrow}^{\downarrow} &     0 \\
    0     &     0 &     0 & \rho_{\rm d}^{\rm d}
  \end{array} \right) \, ,
\end{eqnarray}
since the dot can be either empty ($\chi=0$), occupied with a
spin-up  ($\chi=\uparrow$) or a spin-down ($\chi=\downarrow$)
electron, or doubly occupied ($\chi=\rm d$).
The diagonal elements of the matrix can be interpreted as the
probability to find the dot in the respective state, while the inner
$2 \times 2$ matrix is the $SU(2)$ representation of the average
spin on the dot. All off-diagonal elements connecting different
charge states are prohibited by charge conservation.

For technical reasons it is convenient, to express the density
matrix as vector:  $\rho_{\rm st}=(P_0^0,P_{\uparrow}^{\uparrow},P_{\downarrow}^{\downarrow},
P_{\rm d}^{\rm d},P^{\uparrow}_{\downarrow},P^{\downarrow}_{\uparrow})^T$.
Then the forth-order tensors $\bm  W$'s and $\bm \Pi(\omega)$'s are only
$6\times6$ matrices, see App.~\ref{ws}, and standard computer
implemented matrix operations can be used.
It is worth to point out, that in the vector notation, the
trace for example in Eq.~(\ref{smallw}) is then {\em not} the sum
of all elements of the resulting vector as assumed by
Ref.~\onlinecite{bingdong}, but only the sum of the first four
entries. These elements correspond to the diagonal entries of
the final density matrix. In the notation of
Ref.~\onlinecite{thielmann}, this can be achieved by the vector
$e^T=(1,1,1,1,0,0)$.

The stationary density matrix follows from the master
Eq.~(\ref{master})  $\,0=-i(\varepsilon_\eta - \varepsilon_{\chi})
(\rho_{st})^\chi_{\eta} + \sum_{\chi_1,\eta_1} W^{\chi \chi_1}_{\eta
\eta_1} (\rho_{st})^{\chi_1}_{\eta_1}\,$ under the constraint of
probability normalization $e^T \cdot \rho_{\rm st}=1$. The average
$dc-$current through the system is given by
$I=e/(2\hbar) e^T\cdot {\bm W^I} \cdot\rho_{\rm st}$. In the
low-frequency limit the frequency-dependent propagator
$\bm \Pi(\omega)$ can be constructed from the frequency-dependent free
propagator $\bm \Pi_0(\omega)$ and the frequency-{\em independent}
self energy $\bm W(\omega=0)$. The low-frequency noise is 
then given by the matrix multiplication $S(\omega)=e^2/(2\hbar) e^T\cdot
({\bm W^{II} + \bm W^{I} \bm \Pi(\omega) \bm W^{I}})\cdot
\rho_{\rm st} +(\omega\rightarrow -\omega)$, where the $i0^+$ 
in the denominator of the propagator is dropped, since the term arising from 
the $i0^+$ contribution cancels the delta function in Eq.~(\ref{noise2}).

\section{\label{Results}Results}
\begin{figure}[h!]
\includegraphics[width=0.9\columnwidth]{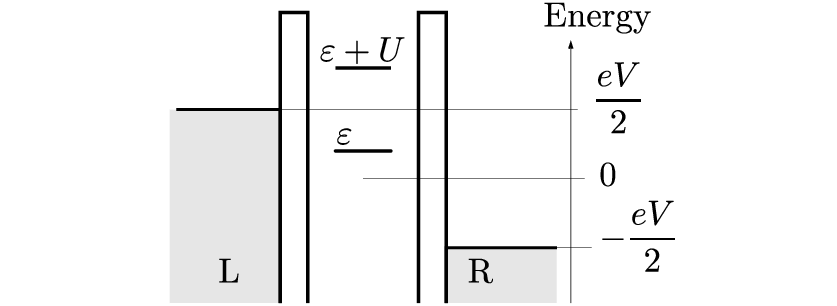}
\caption{\label{fig:dotenergy}
Sketch of different energies involved. Since we 
assume equal tunnel interface capacities, the
voltage  drop on the left and right side is symmetric.}
\end{figure}

In this section, we discuss our results for zero-
and finite-frequency current noise in a quantum dot connected to
ferromagnetic leads with non-collinear magnetizations. The relative
energies of a single-level dot is sketched in
Fig.~\ref{fig:dotenergy}.

We always assume $k_{\rm B}T\gg\Gamma$, and that the single-particle
state is above the equilibrium Fermi energy of the
leads, otherwise higher-order tunnel processes could become
important.\cite{thielmann2,superpoissonian2,assymmetric,Weymann}

\subsection{\label{sec:Zerofrequencynoise}Zero-frequency noise}

\begin{figure}[h]
\includegraphics[width=1.0\columnwidth]{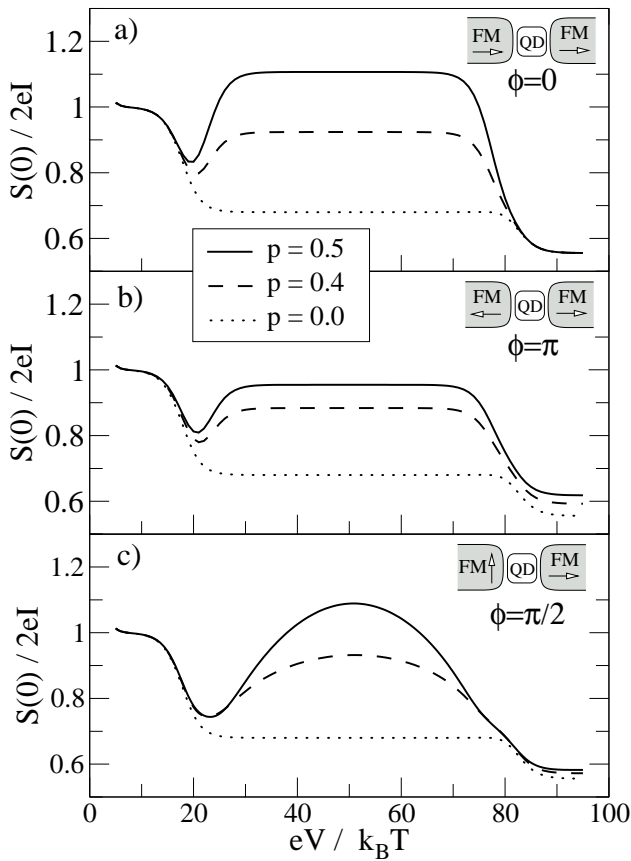}
\caption{\label{zerofrequency}
Zero-frequency current noise through a quantum dot spin valve.
In panel a), the lead magnetizations are aligned parallel, in
panel b) anti-parallel, and in panel c) the lead magnetizations
enclose an angle of $\pi/2$. The different lines correspond to
different values of the lead polarization $p$.
Other parameters are
$\varepsilon=10k_{\rm B}T$, $U=30k_{\rm B}T$,
and $\Gamma_{\rm L}=2\Gamma_{\rm R}$}
\end{figure}

We start our discussion with the zero-frequency noise.
In Fig.~\ref{zerofrequency}, we plot results for
 $F(\omega\rightarrow 0)=S(\omega\rightarrow 0)/(2e I)$,
i.e. the zero-frequency Fano
factor for the quantum dot contacted by ferromagnetic  leads.
In Fig.~\ref{zerofrequency}a), the leads are aligned  \emph{parallel}. For
$eV/2<\varepsilon$, when the dot level is above the lead Fermi
energies, the dot is  predominantly empty, and interaction effects
are negligible leading to a Fano factor of  $1$. In the voltage
window $\varepsilon<eV/2<\varepsilon+U$, when the dot can only be
empty or singly occupied, we can observe super-Poissonian noise due to
dynamical spin  blockade\cite{cottet,superpoissonian2,Bulka1999}
for sufficiently high lead polarization. The
minority spins have a much longer dwell time inside the dot than the
majority spins. In this way, they effectively chop the current leading
to bunches of majority spins. While the current in this regime
$I= 2\Gamma_{\rm L}\Gamma_{\rm R}/(2\Gamma_{\rm L}
+\Gamma_{\rm R})$ does not depend on the polarization $p$ of the
leads, the Fano factor
\begin{eqnarray}\label{w0parallel}
F(0)=
\frac{4\frac{1+p^2}{1-p^2}\Gamma_{\rm L}^2+\Gamma_{\rm R}^2
}{(2\Gamma_{\rm L}+\Gamma_{\rm R})^2}
\end{eqnarray}
even diverges for $p\rightarrow 1$. If the voltage exceeds the
value necessary to occupy the dot with two electrons
($eV/2>\varepsilon+U$), the noise is no longer sensitive to a lead
 polarization.

Also in the case of  \emph{anti-parallel} aligned leads, the Fano factor
rises in the voltage regime $\varepsilon<eV/2<\varepsilon+U$ as seen
in Fig.~\ref{zerofrequency}b). The dot is primarily occupied with an
electron with majority spin of the source lead, i.e. minority spin
for the drain lead, since this spin  has the longest dwell time. If
the electron tunnels to the drain lead, it gets predominantly
replaced by a majority spin of the source lead. For a high enough
lead polarization, only one spin component becomes important. Further
this spin component is strongly coupled to the source lead and weakly
coupling to the drain lead, therefore the Fano factor approaches unity.

If the leads are  \emph{non-collinearly} aligned, for example enclose an angle
$\phi=\pi/2$ in  Fig.~\ref{zerofrequency}c), a qualitatively different
behavior can be observed. Now, the typical Coulomb plateaus are
modulated. This shape arises, since the dot spin starts to precess
around the lead magnetizations. The tunnel coupling between the
ferromagnetic lead $r=$L/R and the dot induces the exchange field contribution\cite{braun1,braun2}
\begin{equation}
\label{exchange}
  {\bm B}_r = p \,\frac{\Gamma_r{\bf \hat n}_r}{\pi\hbar} \int'
  d\omega \left( \frac{f_r(\omega)}{\omega-\varepsilon-U}
    +\frac{1-f_r(\omega)}{\omega-\varepsilon} \right)\,,
\end{equation}
generating an intrinsic spin precession of the dot spin around
the lead magnetizations.  This exchange field automatically
appears in a rigid calculation of the generalized transition
rates $\bm W$.

The intrinsic spin precession due to the exchange field counteracts
the dynamical spin blockade. The exchange coupling to one lead is
maximal, if its Fermi energy coincides with the dot energy levels,
i.e. the coupling to the source lead is maximal at the voltages
$eV/2=\varepsilon$ and  $eV/2=\varepsilon+U$ and changes its sign in
between. Therefore the reduction of the Fano factor is non-monotonic, and
so is the variation of the Coulomb plateaus. It is worth to point
out, that to observe this spin precession mechanism in the
conductance of the device a relatively high spin polarization of the
leads is required. But the noise is much more sensitive to this
effect than the conductance, that a polarization as
expected for Fe, Co, or Ni\cite{pasupathy} is well sufficient.

\begin{figure}[h]
\includegraphics[angle=-90,width=1.0\columnwidth]{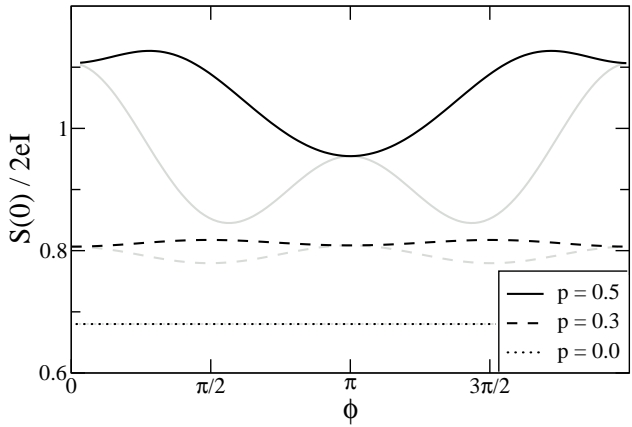}
\caption{\label{noiseplot2}
Fano factor of a quantum dot spin valve as a function of
the angle $\phi$, enclosed by the lead magnetizations.
The black lines are for a bias voltage $eV=50k_{\rm B}T$,
where the exchange field is very weak, while the gray
curves are for $eV=30k_{\rm B}T$, where the exchange
field is more pronounced. Further parameters are
$\varepsilon=10k_{\rm B}T$, $U=30k_{\rm B}T$,
and $\Gamma_{\rm L}=2\Gamma_{\rm R}$ }
\end{figure}

The zero-frequency Fano factor as a function of the angle $\phi$
between the two lead magnetization vectors is plotted in
Fig.~\ref{noiseplot2}. The black lines are for the bias voltage
$eV=50k_{\rm B}T$, where the exchange field influence is weak, while the gray
lines is for the bias voltage $eV=30k_{\rm B}T$. Since both voltages
are within the voltage window allowing only single occupation of the
dot, compare Fig.~\ref{zerofrequency}, the tunnel rates do not
change significantly within this voltage range. Only the exchange
field varies with voltage.  Since the exchange field
suppresses bunching due to spin precession, the black and gray curves split.

For $\phi=0$ and $\phi=\pi$ the accumulated spin is collinearly
aligned with the exchange field, and no spin precession arises.

\subsection{\label{sec:weakmagneticfields}
Finite-frequency noise and weak magnetic fields}
The $dc-$conductance of the quantum-dot spin valve is a direct
measure of the time-averaged spin in the dot. On the other
side, the power spectrum of the current noise can also measure the
time-dependent dynamics of the individual electron spins in the dot. The
spin precesses in the exchange field as well as an external
magnetic field. This gives rise to a signal in the
frequency-dependent noise at the Larmor frequency of the total
field.

By including an external magnetic field in the noise calculation, one
has to  distinguish two different parameter regimes: either the Zeeman
splitting  $\Delta\equiv g\mu_{\rm B}B_{\rm ext}$ is of the same
order of magnitude as the level broadening $\Delta\approx
\Gamma_{\rm L},\Gamma_{\rm R}$, or it significantly exceeds  the
tunnel coupling $\Delta\gg\Gamma_{\rm L},\Gamma_{\rm R}$. In this
section we focus on the first case, while the latter case is
treated in  Sec.~\ref{strongmagneticfields}.

By choosing the spin-quantization axis of the dot subsystem parallel
to the external magnetic field, the magnetic field only induces a
Zeeman splitting of the single-particle level $\varepsilon$ 
in $\varepsilon_\uparrow=\varepsilon+\Delta/2$ and 
$\varepsilon_\downarrow=\varepsilon-\Delta/2$. 
 Since $\Delta \approx \Gamma_{\rm L},\Gamma_{\rm R}$, we can expand 
the $\bm W$'s also in $\Delta$ and keep only the zeroth-order 
terms, since each correction of the self energies would be proportional to 
$\Delta \cdot\Gamma \approx \Gamma^2$. The Zeeman splitting must only
be considered for the free propagator. With Eq.~(\ref{freepropagator}), 
the propagator is then given by
\begin{eqnarray}
  \label{propagatormatrix}
  \bm \Pi_0(\omega)= i \left(
  \begin{array}{cccccc}
    \omega &       0 &        0 &        0 &                 0 &                 0\\
    0        & \omega&        0 &        0 &                 0 &                 0\\
    0        &       0 & \omega &        0 &                 0 &                 0\\
    0        &       0 &        0 & \omega &                 0 &                 0\\
    0        &       0 &        0 &        0 & \omega+\Delta &                 0\\
    0        &       0 &        0 &        0 &                 0 & \omega-\Delta
  \end{array} \right)^{-1} \,,\,\,\,\,
\end{eqnarray}
where we  already dropped the $+i0^+$ in the denominator, and
use the matrix notation as introduced in Sec.~\ref{Technical}.
The two last rows  of this matrix govern the time evolution of
$\rho_{\uparrow}^{\downarrow}$ and  $\rho^{\uparrow}_{\downarrow}$,
representing the spin components transverse to the quantization
axis, i.e. transverse to the applied magnetic field. The change of
the denominator by the Zeeman energy $\Delta$ describes just the
precession movement of the transverse spin component. Since the free
propagator $\bm \Pi_0(\omega)$ is a function of $\Delta$, the Zeeman energy
modifies the full propagator $\bm \Pi(\omega)$ as well as the
(zeroth-order) stationary density matrix  $\bm \rho_{\rm st}$, via the
master Eq.~(\ref{master}).

\begin{figure}[h]
\includegraphics[angle=-90,width=1.0\columnwidth]{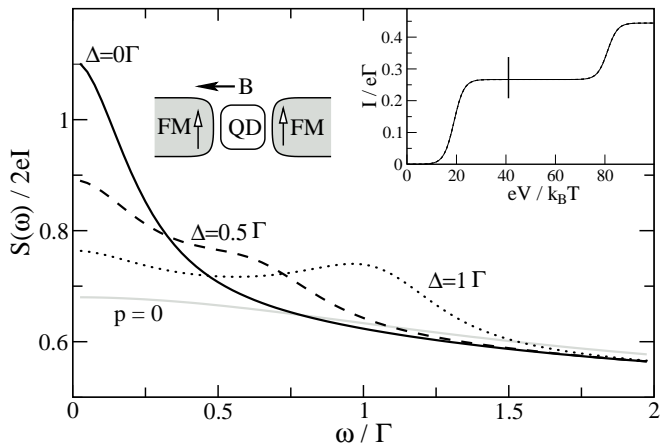}
\caption{\label{parallelleads}
Frequency-dependent Fano factor of a
quantum dot connected to parallel aligned leads for
various perpendicular applied external magnetic fields.
The parameters are
$p=0.5$, $\varepsilon=10k_{\rm B}T$, $U=30k_{\rm B}T$,
$eV=40k_{\rm B}T$ and $\Gamma_{\rm L}=2\Gamma_{\rm R}$.
The inset shows the current bias-voltage characteristic,
which does not depend on the applied magnetic field. }
\end{figure}

\begin{figure}[h]
\includegraphics[angle=-90,width=1.0\columnwidth]{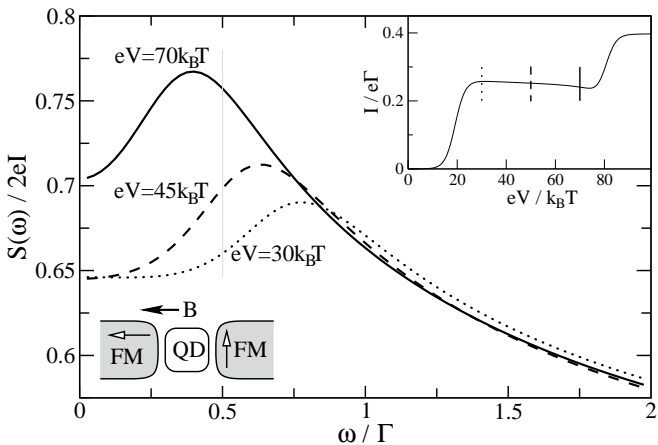}
\caption{\label{noncollinear}
The Fano factor of a quantum dot spin valve
as a function of the noise frequency. The lead 
magnetizations enclose an angle $\phi=\pi/2$ 
and an external magnetic field
$g\mu_{\rm B}B_{\rm ext}=1/2\Gamma$ is applied 
parallel to the source lead magnetization. 
The vertical gray line marks the Larmor frequency 
given by the external magnetic field only.
For the three different bias voltages, 
$eV=30k_{\rm B}T$ (dot-dot-dashed),
$eV=45k_{\rm B}T$ (dashed) and 
$eV=70k_{\rm B}T$ (solid), the strength of the 
exchange field varies, and so does the position of 
the resonance peak. Other system parameters are
$p=0.5$, $\varepsilon=10k_{\rm B}T$, $U=30k_{\rm B}T$,
and $\Gamma_{\rm L}=2\Gamma_{\rm R}$ }
\end{figure}

The numerical results are plotted in Fig.~\ref{parallelleads} and
Fig.~\ref{noncollinear}. In Fig.~\ref{parallelleads} the
magnetizations of the leads are aligned parallel, and a magnetic
field is applied perpendicular to the lead magnetizations. With
parallel aligned leads and equal polarizations in both leads, no
average spin accumulates on the dot, and therefore the
current-voltage characteristic as shown in the inset of
Fig.~\ref{parallelleads}, shows neither magnetoresistance nor
the Hanle effect if a transverse magnetic field is
applied.\cite{braun2}  In contrast to the conductance,
which depends on the average dot spin only, the
frequency-dependent noise is sensitive to the time-dependent
dynamics of the spin on the dot.
Therefore the field-induced spin precession is visible in the
noise power spectrum. For
$B=0$ the Fano factor shows a Lorentzian dependence of the noise
frequency. Thereby the Fano factor exceeds unity due to the bunching
effect, as discussed in Sec.~\ref{sec:Zerofrequencynoise}.

With increasing magnetic field, spin precession lifts the
dynamical spin blockade inside the dot, and the Fano factor
decreases at $\omega\approx0$.

Further a resonance line evolves approximatively at the Larmor
frequency of the applied magnetic field.
The line width of the resonance is given by the damping due to 
tunnel events.  If the dot can only be singly occupied, the 
damping coefficient equals the tunnel-out rate $\Gamma_{\rm R}$, 
if the dot can also be doubly occupied, also tunnel-in events 
contribute.

The deviation of the
resonance line position from the Larmor frequency, one would expect
by considering the applied magnetic field only, is caused by the exchange
interaction. The spin inside the dot precesses in the
total field containing the external magnetic field and the exchange
field.\cite{braun1} Dependent on their relative orientation, the
exchange field can increase or decrease the total field strength.

Since the exchange field is a function of the applied bias voltage, the
resonance peak is shifted by changing the bias voltage.  In
Fig.~\ref{noncollinear}, the finite frequency noise for a quantum 
dot is plotted, where the lead's magnetizations enclose an 
angle $\phi=\pi/2$, i.e. their magnetizations are perpendicular to
each other. Further an external magnetic field is applied parallel 
to the source lead magnetization.  The exchange fields originating 
from the leads has to be added to the external field. 
By varying the bias voltage (without significantly 
changing the transition rates, as indicated in the inset of
Fig.~\ref{noncollinear}) the exchange field varies, and the position 
of the resonance peak is shifted.

\subsection{\label{strongmagneticfields}
Limit of strong magnetic fields}
In this section, we discuss the case of an applied 
magnetic field, where the Zeeman energy 
$\Delta\equiv g\mu_{\rm B}B_{\rm ext}\gg\Gamma_{\rm L},
\Gamma_{\rm R}$ exceeds the tunnel coupling strength. As 
a simplification, we can consider the tunnel rates 
(i.e. the $\bm W$'s) still as independent of $\Delta$ as 
well as of $\omega$. This assumption is justified, if the 
distance between the quantum dot states and lead Fermi 
surfaces well exceeds temperature $k_{\rm B}T$, the Zeeman splitting 
$\Delta$ and the noise frequency $\hbar \omega$.

For a clear analytic expressions, we 
expand the stationary density matrix
in zeroth order in $\Gamma/\Delta$. Further we consider only 
the noise frequency range $\omega=\Delta\pm\Gamma$. In this regime 
the first five diagonal entries of the
free propagator in Eq.~(\ref{propagatormatrix}) can be treated as
zeroth order in $\Gamma$, i.e. their contribution drops out for
the lowest-order noise, only the last entry
$1/(\omega-\Delta)\approx 1/\Gamma$ is kept.
This considerably simplifies the calculation,
since all bunching effects and the exchange field components
perpendicular to the external field can be neglected.

\begin{figure}[h]
\includegraphics[angle=-90,width=1.0\columnwidth]{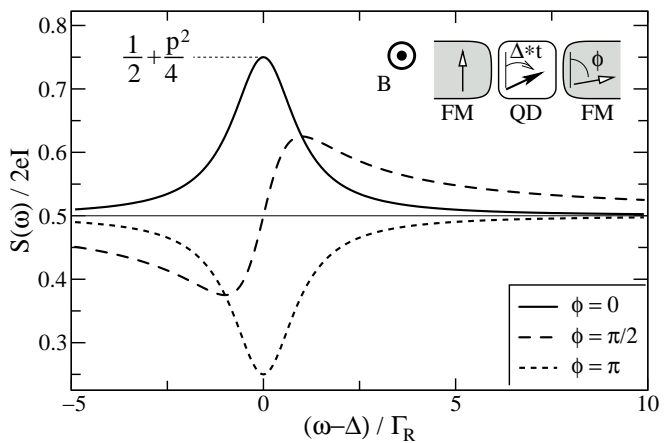}
\caption{\label{finitefrequency}
Fano factor as a function of noise frequency for
different angles $\phi$ in the frequency range
of the Larmor frequency. The applied voltage
does only allow single occupation of the dot.
Other system parameters are as in
Fig.~\ref{noncollinear}.}
\end{figure}

Let us consider a single-level quantum dot with such an applied
voltage, where approximately $f_{\rm L}(\varepsilon)=1$ and
$f_{\rm L}(\varepsilon+U)=f_{\rm R}(\varepsilon)=f_{\rm R}
(\varepsilon+U)=0$, i.e. the applied bias voltage allows only an
empty or singly-occupied dot. For an external applied magnetic field
perpendicular to both lead magnetizations $\bm B_{\rm ext} \perp
\hat{\bm n}_{\rm L},\hat{\bm n}_{\rm R}$ the Fano factor
\begin{eqnarray}
F(\omega)=     \frac{1}{2}+\frac{p^2}{4}\,
\frac{\Gamma_{\rm R}^2 \cos\phi\,\,+\,\,\Gamma_{\rm R} (\omega-\Delta) \sin\phi}{\Gamma_{\rm R}^2+(\omega-\Delta )^2}\,
\end{eqnarray}
shows a resonance signal at the Larmor frequency $\hbar\omega=\Delta$.
By representing the frequency-dependent Fano factor as an integral over time,
\begin{eqnarray}
F(\omega)=\frac{1}{2}+\frac{\Gamma_{\rm R} p^2}{4\hbar}\,
\Re\int_0^{\infty} \!\!\!\!dt\,\, e^{-\Gamma_{\rm R}/\hbar \cdot t}\,\,e^{-i(\Delta/\hbar \cdot t -\phi)}\,\,e^{i\omega t}\, ,\quad
\end{eqnarray}
the discussion of the functional form becomes more transparent. At
$t=0$ an electron tunnels from the source (left) lead in the dot.
This electron decays on average to the drain (right) lead on the
time scale $\hbar/\Gamma_{\rm R}$. During its dwell time the electron
precesses inside the dot with the Larmor frequency $\Delta/\hbar$. This
precession modulates the decay rate, due to magnetoresistance
effects. The tunnel-out event is more likely, if the spin is aligned
parallel to the drain lead magnetization than if anti-parallel
aligned. The phase of this modulation is given by the relative angle
of the lead magnetizations, and the effect can give rise to an
absorption or dispersion line shape, see Fig.~\ref{finitefrequency}.

By shifting the gate voltage such that $f_{\rm L}(\varepsilon)=
f_{\rm L}(\varepsilon+U)=f_{\rm R}(\varepsilon)=1$ and
$f_{\rm R}(\varepsilon+U)=0$, the dot will always be at least 
occupied by one electron. Then the noise shows the same resonance, 
only $\Gamma_{\rm R}$ and $\phi$ must be replaced by
$\Gamma_{\rm L}$ and $-\phi$.

If the leads are aligned parallel, the electron will leave the dot
primary directly after the tunnel-in event, or after one revolution,
i.e. the decay is modulated with a cosine function. If the leads are
aligned perpendicular to each other, then the electron must be
rotated by the angle $\pi/2$ (or $3\pi/2$) before the maximum
probability for the tunneling-out event is reached. The decay is
then modulated by a (minus) sine function.

\begin{figure}[h]
\includegraphics[width=0.9\columnwidth]{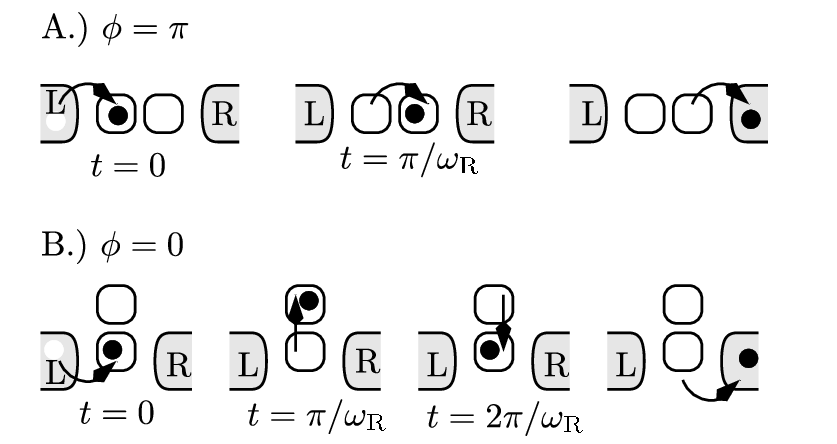}
\caption{\label{ddot}
The double dot analog for the decay phase shift
$\phi=0$ and $\phi=\pi$ of the electrons.}
\end{figure}

The phase dependence of the noise resonance is also predicted for a
double-dot system.\cite{bingdong,gurvitz2} Let us consider two dots 
connected in series, see Fig.~\ref{ddot}A), and an electron from the 
left (source) electrode enters the left dot. Since this is not an 
eigenstate of the isolated double-dot system, the electron coherently 
oscillates between the two dots with the frequency $\omega_{\rm R}$. After
the time $t=\pi/\omega_{\rm R}$, the electron is in the right dot
and can tunnel to the drain lead. This corresponds to the $\phi=\pi$
case resulting a dip in the noise. The realization of the $\phi=0$
case would be a double dot where the left (source) and right (drain)
lead is contacted to the same dot, see Fig.~\ref{ddot}B). Here the
electron must stay a multiple of $2\pi/\omega_{\rm R}$ inside the
double dot to tunnel to the drain lead, giving a peak in the
frequency noise spectrum. Other values of $\phi$ have no double-dot-system 
analogon.

\subsection{\label{spinrelaxation}Influence of spin relaxation}
The density matrix approach offers a way to phenomenologically
include spin relaxation by supplementing the matrix $\bm W$ by

\begin{eqnarray}
  \label{matrixdecay}
  \bm W^\prime= \bm W +\hbar\left(
  \begin{array}{cccccc}
    0        &       0 &        0 &        0 &        0 &        0 \\
    0        & -\frac{1}{T_1} &  +\frac{1}{T_1} &        0 &        0 &        0 \\
    0        & +\frac{1}{T_1} &  -\frac{1}{T_1} &        0 &        0 &        0 \\
    0        &       0 &        0 &        0 &        0 &        0 \\
    0        &       0 &        0 &        0 &  -\frac{1}{T_2} &        0 \\
    0        &       0 &        0 &        0 &        0 &  -\frac{1}{T_2}
  \end{array} \right) \, .\,
\end{eqnarray}
The entries in the lower right corner of Eq.~(\ref{matrixdecay})
describe the exponential decay of the transverse spin components
on the time scale $T_2$,
and the block in the upper left corner describes an equilibration
of the occupation probability for spin up and down.
If one define the average spin vector on the quantum dot by
$\bm S=( \rho^\uparrow_\downarrow+\rho^\downarrow_\uparrow,
i\rho^\uparrow_\downarrow-i\rho_\uparrow^\downarrow,
\rho_\uparrow^\uparrow-\rho^\downarrow_\downarrow)/2$ the master
Eq.~(\ref{master}) becomes a Bloch equation.\cite{braun1} 
The new term in Eq.~(\ref{matrixdecay}) introduces an additional
exponential decay term in this Bloch equation.
In the limit of weak Zeeman splitting as discussed throughout
the paper, $T_1$ and $T_2$ become equal, and
$\bm W^\prime$ includes an isotropic exponential damping
of the spin on the dot. Thereby the master equation describing
the change of the probability $\partial_t (P_\uparrow^\uparrow +
P_\downarrow^\downarrow)$ for single occupation is not affected
by this relaxation term.

\begin{figure}[h]
\includegraphics[angle=-90,width=1.0\columnwidth]{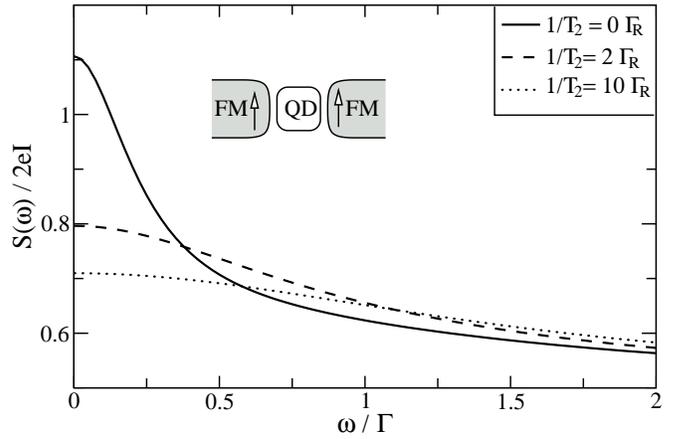}
\caption{\label{damping}
Frequency dependence of the Fano factor if the leads are
aligned parallel. With increasing spin relaxation, the spin
blockade and therefore the bunching effect is reduced. Other
system parameters are as in Fig.~\ref{parallelleads}.}
\end{figure}

The modified rate matrix $\bm W^\prime$ enters the noise calculation via
the calculation of  the stationary density matrix and via the
propagator $\bm \Pi(\omega)$. The numerical solution for the case of 
parallel aligned lead magnetizations is plotted in 
Fig.~\ref{damping}.

With increasing  the spin decoherence, the spin related effects
decrease, which is the expected behavior for spin-decoherence.
To completely suppress the spin related effects the 
spin life time must significantly exceed the inverse tunnel
coupling, i.e. the spin related effects are not very fragile 
against spin-decoherence.

Several articles\cite{RudzinskiBarnas, SouzaEguesJauho,bingdong} try 
to model spin relaxation by the Hamiltonian $H_{\rm rel}= R c^\dag_
\uparrow c_\downarrow+R^\star c^\dag_\downarrow c_\uparrow \,,$ 
which is from the physical point of view dissatisfying, since it
 does not describe incoherent relaxation processes but coherent 
precession in a transverse magnetic field.\cite{QM1} 
This ansatz leads to a completely different 
behavior of the frequency-dependent current noise. Instead of a 
suppression of all spin-related effects with increasing the 
parameter $R$, as expected for spin relaxation, an external field
generates a resonance line. With increasing 
the field strength, this line just shifts to higher and higher 
frequencies, but does not vanish.

\section{\label{conclusions}Conclusions}

By contacting a quantum dot to ferromagnetic leads, the transport 
characteristic through the device crucially depends on the 
quantum-dot spin. In this paper we discussed the 
influence of the spin precession of the dot electron in the tunnel-induced 
exchange field and an applied external magnetic field.
While the conductance depends only on the time-average dot spin, 
the current-current correlation function is sensitive to its 
time-dependent evolution.

In the zero-frequency limit, the spin precession lift the dynamical spin 
blockade, and therefore reduce the zero-frequency noise.
At the Larmor frequency, corresponding to the sum 
of exchange and applied field, the single-spin precession leads to a 
resonance in the frequency dependent current-current correlation function.
Responsible for the resonance is the tunnel-out process of a dot electron 
to the drain lead. Due to magnetoresistance, the tunnel probability depend 
on the relative angle of dot spin and drain magnetization. Therefore the 
spin precession leads to an oscillation of the tunnel probability, visible 
in the current-current correlation function.
The shape of the resonance in the current-current correlation can
either have an absorption or dispersion lineshape, depending on the 
relative angle between the lead magnetizations.

Finally, we show how to properly include spin decoherence, and discuss 
why modelling spin relaxation by an external field transverse to the spin 
quantization axis, as done sometimes in the literature, is unsatisfying.

\begin{acknowledgments}
 We thank J. Barnas, C. Flindt, M. Hettler, B. Kubala, S. Maekawa,
G. Sch\"on, A. Thielmann, and D. Urban for discussions. This
work was supported by the Deutsche Forschungsgemeinschaft under the Center 
for Functional Nanostructures, through SFB491 and GRK726, by the EC under 
the Spintronics Network RTN2-2001-00440, and the Center of Excellence for 
Magnetic and Molecular Materials for Future Electronics G5MA-CT-2002-04049 as
well as Project PBZ/KBN/ 044/P03/2001.
\end{acknowledgments}

\appendix
\section{\label{ws} Generalized transition rates}
The generalized transition matrix $\bm W$ is given by the solution
of the self energy diagrams up to linear order in the coupling
strength $\Gamma$. We have choosen the quantization axis perpendicular 
to both lead magnetizations, and the $x-$axis symmetric with 
respect to the magnetizations. Arranged in the matrix notation
introduced in Sec.~\ref{Technical} we get
\begin{eqnarray}
  \bm W\bigl|_{\omega=0}=\Gamma_{\rm L}\, {\bf A}_{\rm L}+({\rm L}\rightarrow {\rm R})\,,\qquad
\end{eqnarray}
with the matrix $\bm A_{\rm L}$ given by
\begin{widetext}
\begin{eqnarray}
  \label{Wmatrix}
\left(
  \begin{array}{cccccc}
    -2f^+_{\rm L}(\varepsilon) & f^-_{\rm L}(\varepsilon) &   f^-_{\rm L}(\varepsilon) & 0 &
     pf^-_{\rm L}(\varepsilon)\,e^{i\phi_{\rm L}} & pf^-_{\rm L}(\varepsilon)e^{-i\phi_{\rm L}} \\

    f^+_{\rm L}(\varepsilon) & -y_{\rm L} &   0 & f^-_{\rm L}(\varepsilon+U) &
     -\frac{p}{2}(x_{\rm L}-iB_{\rm L})e^{i\phi_{\rm L}} & -\frac{p}{2}(x_{\rm L}+iB_{\rm L})e^{-i\phi_{\rm L}} \\

    f^+_{\rm L}(\varepsilon) & 0 & -y_{\rm L} &   f^-_{\rm L}(\varepsilon+U) &
     -\frac{p}{2}(x_{\rm L}+iB_{\rm L})e^{i\phi_{\rm L}} & -\frac{p}{2}(x_{\rm L}-iB_{\rm L})e^{-i\phi_{\rm L}} \\

    0 & f^+_{\rm L}(\varepsilon+U) &   f^+_{\rm L}(\varepsilon+U) & -2f^-_{\rm L}(\varepsilon+U) &
     -pf^+_{\rm L}(\varepsilon+U)\,e^{i\phi_{\rm L}} & -pf^+_{\rm L}(\varepsilon+U)e^{-i\phi_{\rm L}} \\

    pf^+_{\rm L}(\varepsilon)e^{-i\phi_{\rm L}} & -\frac{p}{2}(x_{\rm L}-iB_{\rm L})e^{-i\phi_{\rm L}} & -\frac{p}{2}(x_{\rm L}+iB_{\rm L})e^{-i\phi_{\rm L}}& pf^-_{\rm L}(\varepsilon)e^{-i\phi_{\rm L}} &-y_{\rm L} &  0 \\
    pf^+_{\rm L}(\varepsilon)e^{+i\phi_{\rm L}} & -\frac{p}{2}(x_{\rm L}+iB_{\rm L})e^{i\phi_{\rm L}}& -\frac{p}{2}(x_{\rm L}-iB_{\rm L})e^{i\phi_{\rm L}}& pf^-_{\rm L}(\varepsilon)e^{-i\phi_{\rm L}}& 0 & -y_{\rm L}\\
  \end{array} \right)\,. \nonumber
\end{eqnarray}
\end{widetext}
The angle $\phi=2\phi_{\rm L}=-2\phi_{\rm R}$ is the angle enclosed by the lead
magnetizations. The leads are characterized by the Fermi functions
$f_r^+(\omega)$ and $f^-_r=1-f^+_r$. For shorter notation we further
introduced $x_{\rm L}=f_{\rm L}^-(\varepsilon)-f_{\rm L}^+(\varepsilon+U)$,
$y_{\rm L}=f_{\rm L}^-(\varepsilon)+f_{\rm L}^+(\varepsilon+U)$, and the exchange field strength 
$B_r=|\bm B_r|/\Gamma_r$, see Eq.~(\ref{exchange}).

\end{document}